\documentclass[12pt,prl,floatfix, showkeys]{revtex4}

\usepackage{epsfig}
\usepackage{amsmath}
\usepackage{amssymb}
\usepackage{graphicx}
\usepackage{latexsym}
\usepackage{rotating}
\usepackage{revsymb}
\usepackage{mathrsfs}

\begin{document}

\renewcommand{\thefootnote}{\fnsymbol{footnote}} 
\renewcommand{\theequation}{\arabic{section}.\arabic{equation}}
\newcommand\degrees[1]{\ensuremath{#1^\circ}}

\title{Predicting the Structure of Alloys using Genetic Algorithms}

\author{Chris E. Mohn and Svein St{\o}len}
% \email[Corresponding author:]{chrism@kjemi.uio.no}
\affiliation{Department of Chemistry and Centre for Materials Science
 and Nanotechnology, University of Oslo, P.O.Box 1033 Blindern, N-0315 Oslo, Norway}
\author{Walter Kob}
\affiliation{Laboratoire des Collo\"{i}des, Verres et Nanomat\'{e}riaux, UMR 5587, Universit\'{e} 
Montpellier 2 - CNRS, 34095 Montpellier, France  }

\begin{abstract}

We discuss a novel genetic algorithm that can be used to find
global minima on the potential energy surface of disordered ceramics
and alloys using a real-space symmetry adapted crossover. Due to a high
number of symmetrically equivalent solutions of many alloys a conventional genetic
algorithms using reasonable population sizes are unable
to locate the global minima for even the smallest systems.  
We demonstrate the  superior performance of the use of symmetry adapted crossover
by the comparison of that of a conventional GA for
finding global minima of two  binary  Ising-type alloys that either 
order or phase separate at low temperature. 
Comparison  of different representations and
crossover operations show that the use of real-space crossover
outperforms crossover operators working on binary representations 
by several orders of magnitude.  

\end{abstract}

% insert suggested PACS numbers in braces on next line \pacs{}

\keywords{Genetic algorithm, atomistic simulation structure prediction, Alloys, Minerals}

\maketitle

\section{Introduction} 

Atomistic processes such as the self-assembly of nano-particles and the
growth of thin films have become extremely important for the development
of new micro and nano devices and therefore have a huge potential in a
vast number of disciplines~\cite{Redl:2003, Shevchenko:2006, Switzer:1999}.  
Understanding at the atomistic level the driving force for these processes
is an essential aspect in order to direct future experimental research
but present an enormous challenge to theory. The complexity of these
processes are often too extensive for an exhaustive brute-force enumeration and
would require an inaccessible amount of computational resources.

Heuristics  have recently challenged conventional
 methods such as Monte Carlo for  understanding complex processes at the
atomistic level and particular promising approaches include evolutionary
and genetic algorithms (GA)~\cite{Holland:1975, Goldberg:1989, Chakraborti:2004,Paszkowicz:2009}.  Powerful
examples of the applications of atomistic GA are the prediction of crystal
structures~\cite{Woodley:1999, Woodley:2008} and the geometry of 
nano-clusters~\cite{Hartke:2004, Johnston:2003, Woodley:2009, Dugan:2009}, 
reconstruction and structure of surfaces~\cite{Chuang:2004, Ciabanu:2009, Mohn:2009gen},
modeling grain boundaries~\cite{Smith:1992, Zhang:2009}, spin
glasses~\cite{Pal:1996} and disordered bulk materials~\cite{Smith:1992,
Mohn:2005gen, Mohn:2009gen, Kim:2005, Chen:2007}.  Inspired by Darwin's
theory of evolution, key operators in the GA are ``selection'' and
``crossover''. The problem of interest is
usually reformulated such that a microstate is mapped on the form of a binary
string. Parents with high fitness (i.e.~low energy) are preferentially
selected for (selection), and offspring are formed by first cutting the
parent strings at certain bit-positions and then combining complimentary
strings of each parents (crossover). Alternatively the crossover
can be made directly in real space (using the original variables)
by combining (often large) ``clusters'' from each parents to form
offspring(s)~\cite{Deaven:1995}. Compared to that of uniform crossover
(i.e.~random bits are taken from each parents) real space operators are more
frequently used within the field of atomistic design since they have
the advantage of preserving local order/structure.

Inspired by the success of GA in the fields mentioned above, we have
recently developed~\cite{Mohn:2009gen} a novel genetic algorithm for
finding configurations with target properties of substitutionally
disordered materials (crystalline materials where different atoms are
distributed over a set of distinct sites) such as alloys (e.g.~AuPt),
ceramics (e.g.~MgO-MnO) and minerals (e.g.~Pyrope-Grossular
 Mg$_3$Al$_2$Si$_3$O$_{12}$-Ca$_3$Al$_2$Si$_3$O$_{12}$). Understanding
local order within substitutionally disordered materials are important within a 
range of field such as the development of new alloys 
with superior mechanical and electrical properties and also for 
understanding geochemical processes within the 
earths interior. 

\begin{figure}[th]
\begin{center}
    \includegraphics[scale=0.8]{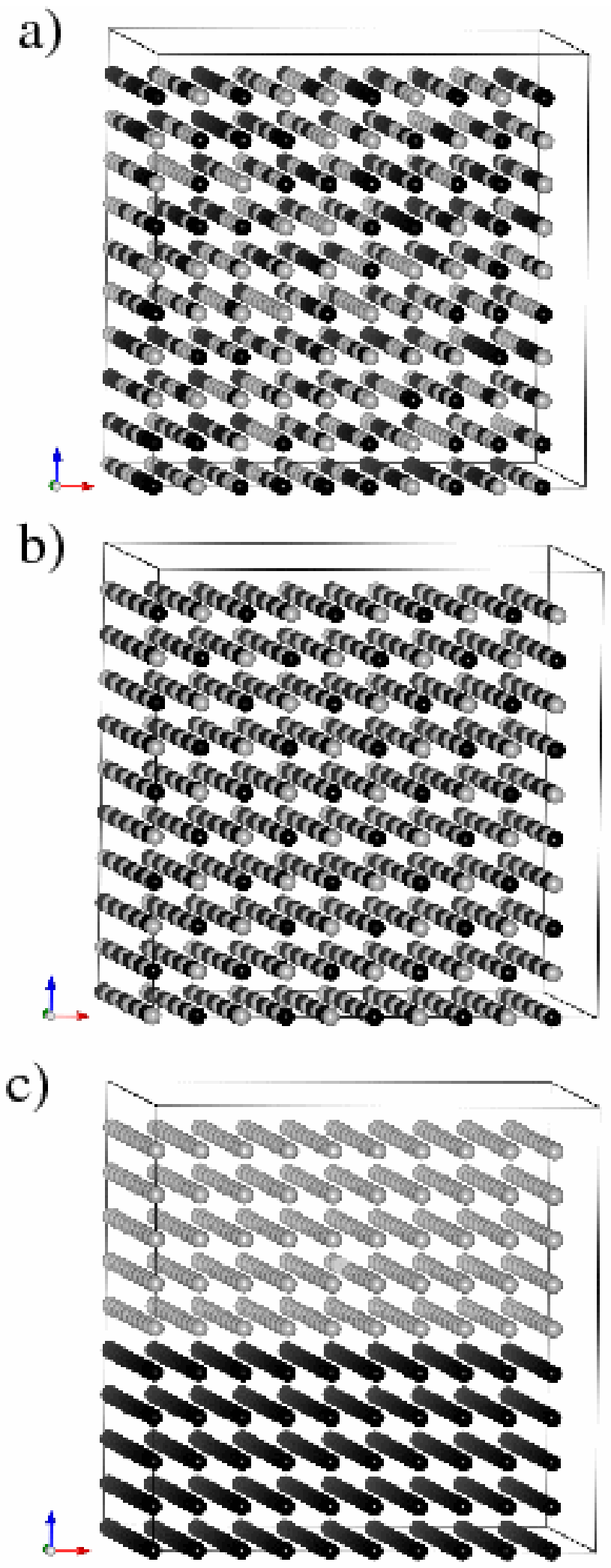}
\end{center}
\caption{An ``AB'' alloy where an equal amount of species ``A'' and
species ``B'' are distributed over the sites of a cubic
lattice within a simulation box consisting of 1000 atoms. a) represent
a ``random'' (high-temperature) alloy, b) is an example of an alloy that
orders at low temperature, whereas c) is a low temperature representation
of an alloy that phase separates.}
\label{alloys}
\end{figure}

A generic high temperature binary ``AB'' alloy is visualised in
Fig.~\ref{alloys}a  where species  ``A'' and ``B'' are distributed at
random on the underlying lattice.  Problems of particular interest are
whether these alloys at low temperature order (see Fig.~\ref{alloys}b)
or phase-separate (see Fig.~\ref{alloys}c) , and if they order we
would like to know how the atoms tend to arrange themselves.  Thus we
are searching for configurations with the lowest energy, which in
general is a hard problem and unit cells with more than 10$^3$ atoms is
typically required. For example, distributing 500 atoms of type ``A''
and 500 atoms of type ``B'' within a simulation box having 1000
lattice positions gives a total of $\approx 10^{300}$ 
distinct arrangements!

Although these problems are in principle suitable for GA, one of the
major problems that hamper conventional GA to find global minima of
alloys is the presence of a large number of symmetrically equivalent (low
energy) configurations, which in general are far away from one-another
in configuration space~\cite{VanHoyweghen:2002}. This can be examplified
using a simple one-dimensional 100 atoms Ising-alloy where
nearest unlike neighbouring atoms repel one another and nearest like neighbouring atoms
attract one another. We apply periodic boundary conditions such that the 
first atom in the string is bonded to the last atom in the string. This
system has the symmetry of a 100-sided polygon and transforms according to the irreducible  
representations of the dihedral group, $D_{100}$. There are 100 distinct (symmetrically equivalent) global minima, as for example:

\begin{figure}[th]
\begin{center}
    \includegraphics[scale=0.4]{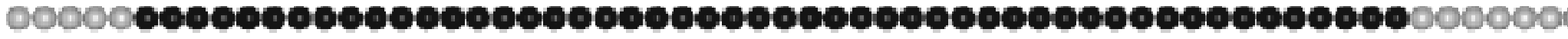}
\end{center}
\end{figure}

\noindent
(Here the grey circles stand for one type of atom and the black ones
for the other type.)

\begin{figure}[th]
\begin{center}
    \includegraphics[scale=0.8]{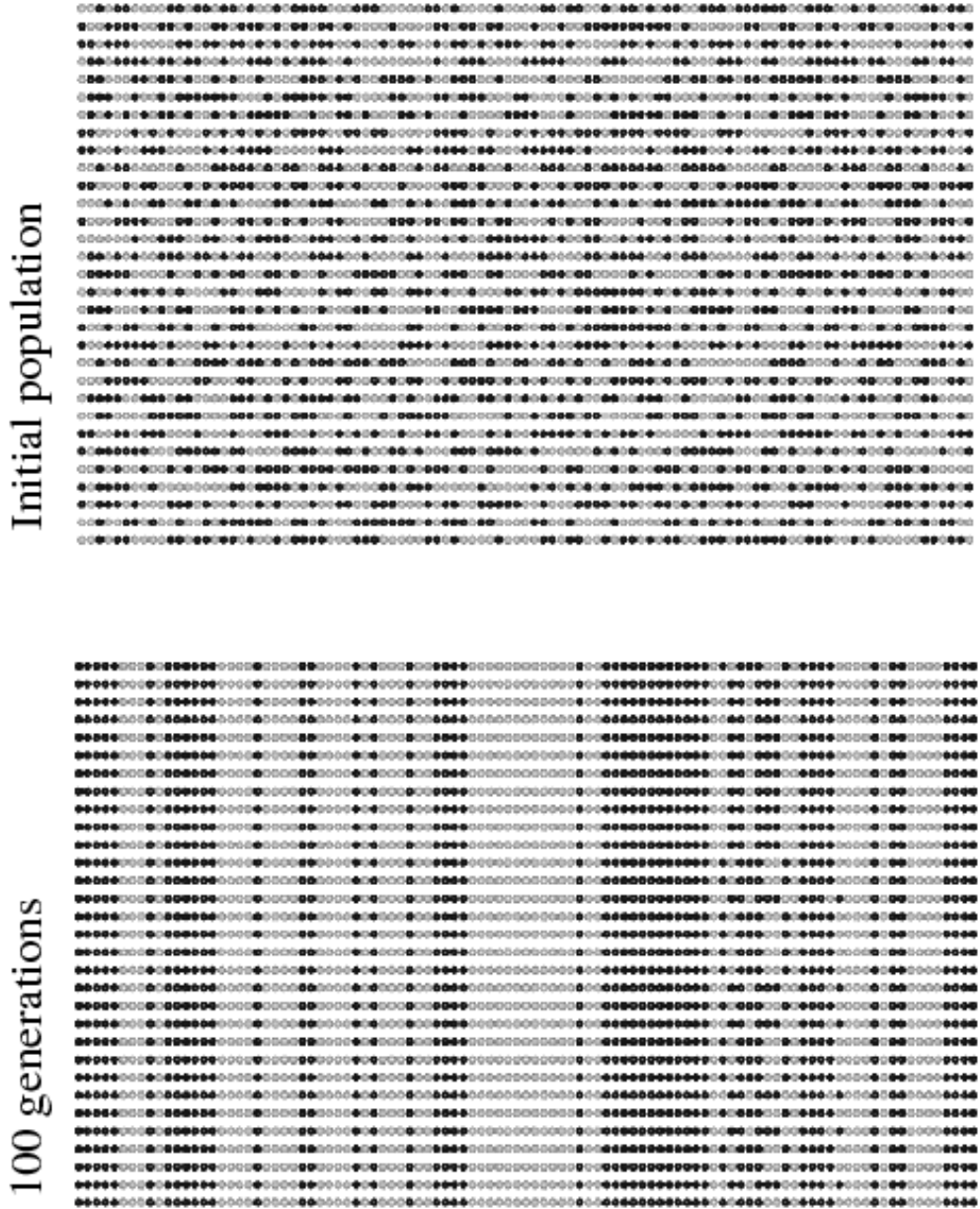}
\end{center}
\caption{The evolution at two stages showing the initial population
(top panel) and the final population (bottom panel) using a conventional
genetic algorithm applied to a periodic one-dimensional Ising model
having 100 spins with a fixed 50:50 composition of species A (black)
and B (grey). A steady state GA with a population of 30 members were
used. The child replaces the worst fit member in the population and no
mutations were applied.}
\label{evolconv}
\end{figure}

Despite the simplicity of this problem, a conventional GA fails to find
one of the global minima even with huge initial populations, because
the population does not contain sufficient diversity to ``differ''
between the building blocks of the low energy minima and the global
minima.  The high order building blocks (i.e.~good local structures,
also called schemata) of a given global minimum are very similar to
those present in a neighbouring low energy minima since the schemata of
the global minimum and the neighbouring low energy minima all contain
high order building blocks such as $\bullet\bullet$$\bullet\bullet $,
$\circ\circ$$\circ\circ $ located at the {\it same} bit positions.
However, since the symmetrically equivalent global minima are 
often {\it far away} from one another in the configurational space, the crossing of
two good minima will typically produce equally good or poorer solutions.
As a consequence the driving force in the direction of a {\it single}
global minimum is very weak and the algorithm is not able to find the
optimal structure.  The premature convergence is manifested as small
(local) clusters of good solutions which cannot recombine to form higher
order building blocks of better solutions.  This is visualised
in Fig.~\ref{evolconv} where we show two stages of a population of 30
members of the simple 1D Ising model introduced above.
One possibility to cure this problem is to introduce random mutations
in the population pool. However, in that case the GA reduces to that of a
random walk with no efficient crossover and thus becomes unsuitable for
finding global minima for alloys in which other methods such as Monte
Carlo algorithms perform better.

We have recently shown~\cite{Mohn:2009gen} that this premature convergence
can be avoided using a real-space {\it symmetry adapted} crossover
(SAC) by replacing the offspring with a randomly chosen symmetrically
equivalent offspring (they have the same fitness!). For example, if
the child looks like

\begin{figure}[th]
\begin{center}
    \includegraphics[scale=0.4]{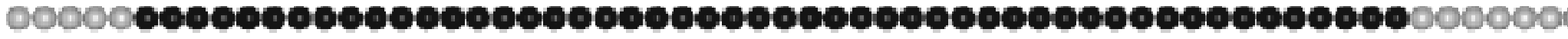}
\end{center}
\end{figure}

we now apply a random translation operation of D$_{100}$ 
by translating all atoms, for example 6 bits to the left. This translation operation 
creates the following symmetrically equivalent child

\begin{figure}[th]
\begin{center}
    \includegraphics[scale=0.4]{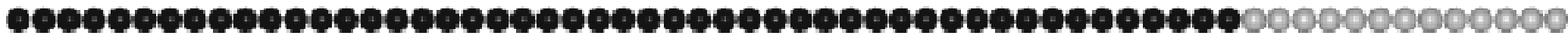}
\end{center}
\end{figure}.

This child, which has the same fitness as the original child, 
is now added to the population rather than the original child.

\begin{figure}[th]
\begin{center}
    \includegraphics[scale=0.7]{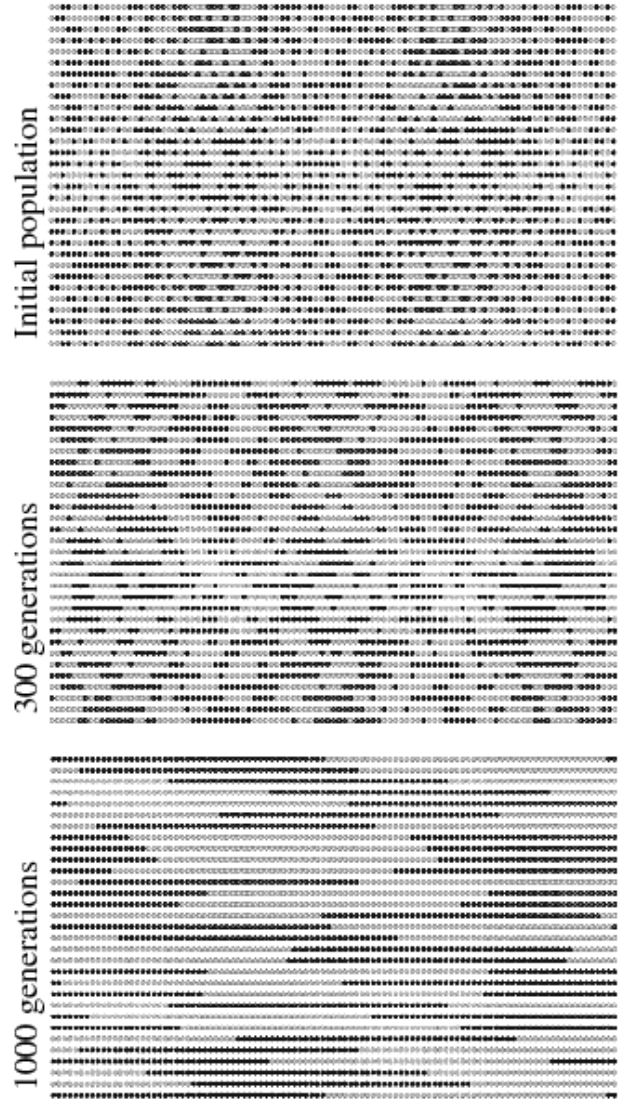}
\end{center}
\caption{The evolution at three stages showing the initial population (top figure), 
the population after 300 steps (middle figure) and the final population (bottom figure) using SAC
genetic algorithm applied to a periodic 1D 100 spin Ising model with a
fixed 50:50 composition of species A (black) and B (grey). A steady state GA 
with a population of 30 members were used. The child
replaces the worst fit member in the population and no mutations were applied.}\label{evolsac}
\end{figure}

The huge advantage of applying such symmetry-adapted mating operations 
is because it allows one to carry out large swaps
in configuration space without destroying the schemata.
This enables the building blocks of good solution to combine to form higher
order building blocks of even better solutions. The evolution of the
periodic 1D Ising-alloy discussed above using symmetry adapted crossover
is shown in Fig.~\ref{evolsac}~\cite{Mohn:2009gen}.  As can be seen,
cluster of ``good'' solutions (Fig.~\ref{evolsac} after 300 generations)  are found at 
{\it all} regions in configurational space, providing a driving force in the
direction of the global minima (Fig.~\ref{evolsac} after 1000 generations).

In the present work we demonstrate the performance of GA in conjunction
with a symmetry adapted crossover applied to two 3D Ising-type problems
with nearest neighbour interactions, representing alloys that either
order or phase separate at low temperature.  These are important benchmark
tests since the global minima are trivally known.

\section{Theory}

\begin{figure}[th]
\begin{center}
    \includegraphics[scale=0.55]{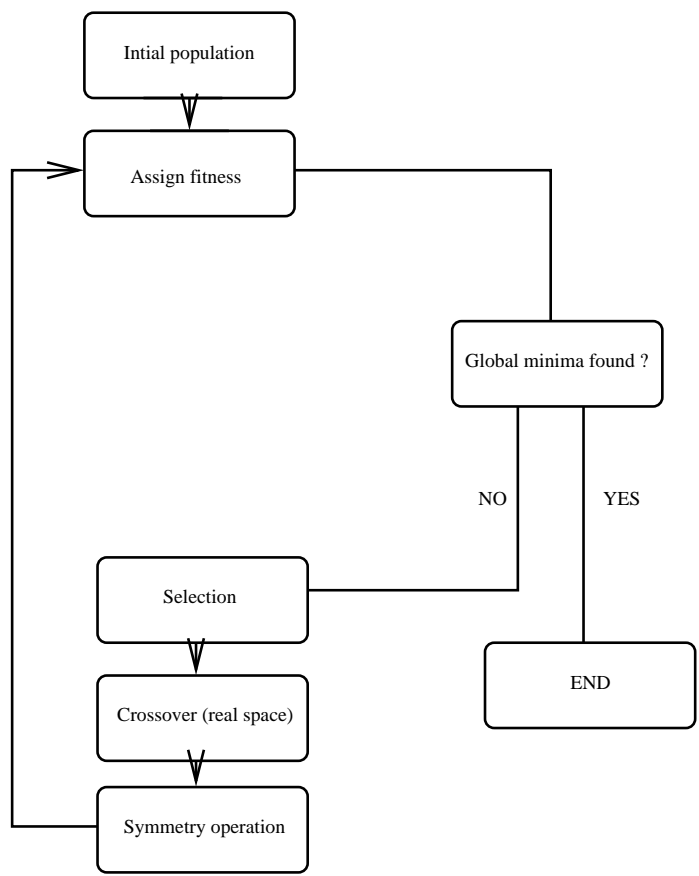}
\end{center}
\caption{Implementation of the GA.}\label{float}
\end{figure}

An implementation of our GA for the study of disordered alloys is 
shown in Fig.~\ref{float}. Different symmetry adapted real-space crossover operations 
have been implemented for all cubic (face centred, body centred and primitive), 
tetragonal (body centred and primitive) and 
orthorombic (face-centred, singe face-centred, body-centred and 
primitive) lattices. In addition, conventional binary operations 
such as two-point and uniform crossover are  implemented but most
attention has been given to different real space crossover operations. 
In this work we use periodic planar cuts which are carried out by choosing two random
points r$_1$ and r$_2$  ($ r_2 > r_1$) along a randomly chozen axis~$\chi$. The child are formed by taking
the atoms from parent 1 with~$\chi$~coordinates, $ r_1 < r_{\text{parent~1}} < r_2 $ and the complimentary atoms from 
parent 2. We have also implemented 
real space crossover operations where various shaped ``clusters'' from one parent
are inserted within the other parent.

The energy we use for studying 3D Ising alloys was originally introduced for the study of
ferromagnetism in 1925 by Ising~\cite{Ising:1925} and is given by

\begin{equation}
  E = -\frac{1}{2}\sum_{<ij>}J\sigma_i \sigma _j.
\label{eq_ising} 
\end{equation}

\noindent
Here $\sigma_i$, are  either $+1$ if the atoms are of type ``A'' or
$-1$ if the atoms are of type ``B'' and $J$ is the interaction energy
between the species. For simplicity, the summation is carried over
the nearest neighbours only.  If $J<0$, bonds between unlike atoms
contribute $-|J|$ to the total energy and bonds between like pairs
contribute $+|J|$ to the total energy which promote a high degree
of mixing between the difference species and the ground state is an
alternating sequence of $+1$ and $-1$ spins/atoms.  If $J>0$, unlike atoms
repel one-another promoting phase-separations at low temperature. The
summation in equation~\ref{eq_ising} is {\it constrained} to fix the
overall composition during the evolution.In this work we investigate two 
such 3D Ising alloys in which the atoms are distributed on
a cubic lattice.  

In all runs an intial population 
was selected at random. We apply a tournament selection, in 
which the candidates with the highest fitness (lowest energy) among two 
randomly chosen pairs are selected
as parents. We use a steady state genetic algorithm
in which at each generation one of the children replaces the worst fit
member in the population. 

\section{Results}

\begin{figure}[th]
\begin{center}
    \includegraphics[scale=0.55]{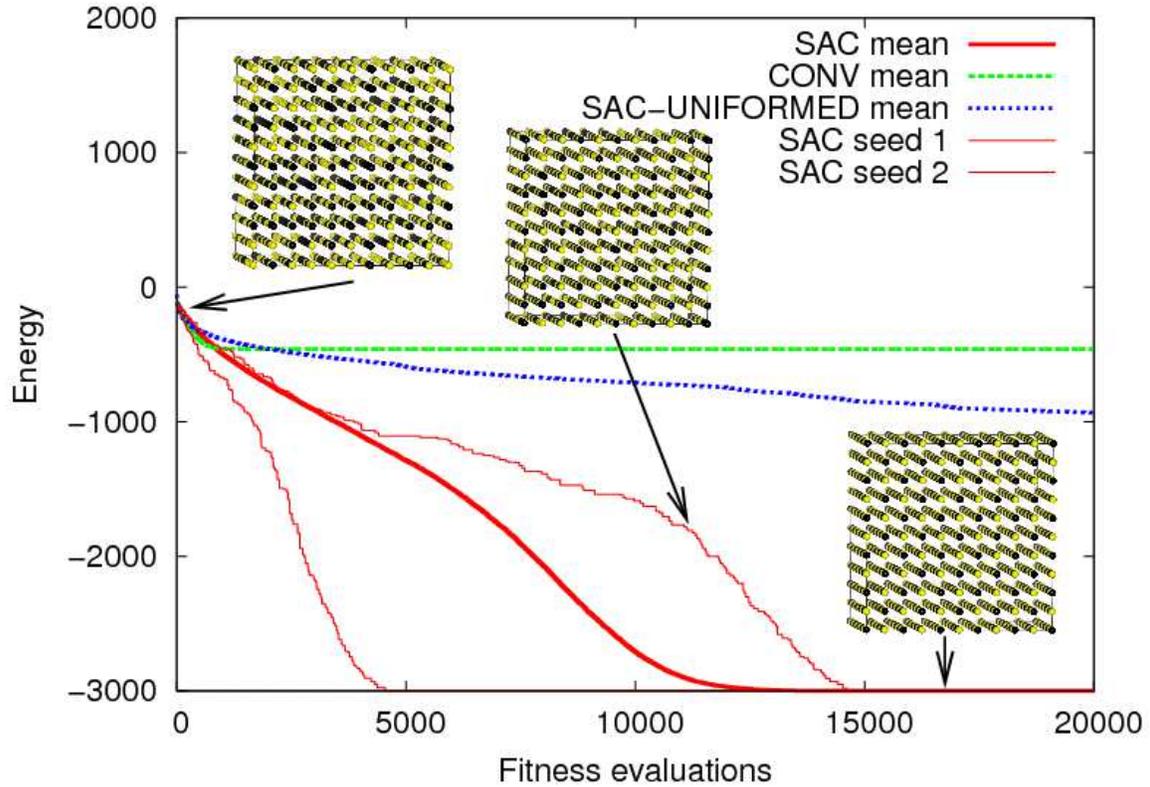}
\end{center}
\caption{Progression plots of the GA of a periodic primitive cubic alloy
with 1000 atoms (a 50:50 sample of species ``A'' and ``B'' ) with $J =
-1$.  The mean value based on 100 samples (bold full line) along with two seeds are shown
using real space SAC (thin red lines) and a conventional (CONV) GA (green
dashed lines). The mean value based on 100 samples using a uniformed SAC
is shown as a blue dotted line.  The lowest energy structures at three
stages is shown. Further computational details are given in the text.}
\label{prog1}
\end{figure}

Results from GA calculations of a primitive cubic periodic 1000-atoms
Ising alloy with $J = -1$  and a fixed 50:50 composition of species
A and B are shown in Fig.~\ref{prog1}.  Simulations were carried out
both with and without the use of symmetry adapted crossover (bold
solid and dashed lines, respectively) and a population size of 50
members. We also show results where a uniformed SAC, described below,
was used (bold dotted line).  100 independent runs were carried out,
and the mean value of the best energies at each step is displayed in
Fig.~\ref{prog1}. We also include two individual runs, marked as
seed 1 and seed 2  (thin lines) using real space SAC. Note that the ground state 
has an energy of 3 times the number of atoms, i.e.~here 3000.

As can be seen from the figure, the use of a real space symmetry-adapted
crossover clearly outperforms the conventional GA which fails in all
attempts to find the global minimum.  Even with huge population sizes
($>10 000$ members) the conventional genetic algorithm is unable to find a
global minimum.  By contrast, we find that the use of SAC in combintation 
with real space crossover is very robust
even with small populations.  In all 100 real space SAC optimisations less than
15000  fitness evaluations were required to find the global minium.
Even population sizes consisting of only 10 members were sufficient
in order to locate the global minima in {\it all} 100 simulations.
On average, only 2000 steps were needed to find the global minima with
a population size of 10 members, whereas 10000 steps were needed when
the size of the population was 50. In the former case the global minima
were found in all simulations in less than 6000 steps whereas 96\% were
found in less than 3000 steps.  These results are highly encouraging
if one bears in mind the extreme large number of possible solutions.
A uniformed crossover (by taking random atoms from parent 1 and the
complimentary set of atoms from parents two) together with SAC is also
able to find the global minima, but the evolution is very slow as seen in
Fig.~\ref{prog1} and more than $10^6$ steps are, on average, needed to
find a global minimum. The slow convergence is easily explained because
many bonds are broken when a uniformed crossover is used and potentially
good building blocks are frequently destroyed.

We have also carried out tests using simple mutations by, e.g., exchanging
a random pair of different atoms but, in fact, we found that the evolution
is slowed down when mutations are applied.

\begin{figure}[th]
\begin{center}
    \includegraphics[scale=0.55]{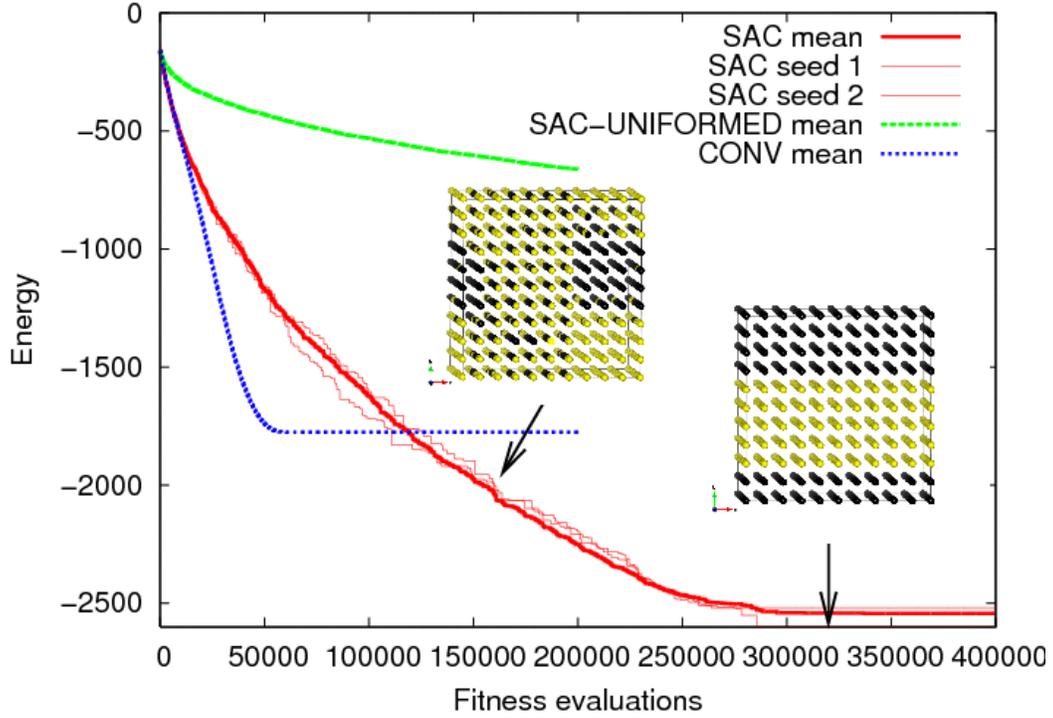}
\end{center}
\caption{Progression plots of the GA for a periodic primitive cubic
alloy with 1000 atoms (a 50:50 sample of species ``A'' and ``B'' ) with
$J = 1$.  The mean value (bold line) is based on 100 samples. Also shown
are two runs with two seeds using real space SAC (thin red lines) and
a conventional (CONV) GA (green dotted lines).  The mean value based
on 100 sample using a uniformed SAC is shown as a bold blue dotted
line.  The lowest energy structures at three stages is shown. Further
computational details are given in the text.}
\label{prog2}
\end{figure}

In Fig.~\ref{prog2} we plot results for the case $J = 1$.  For this
value the configurations with the lowest energy is the one in which the
atomic species are completely segregated, mimicking systems that phase
separate at low temperature as shown in Fig.~\ref{alloys}c.  Again,
1000 atoms were placed on a primitive cubic simulation box and a fixed
50:50 composition of species A and B was studied, and periodic boundary
conditions were imposed (i.e. the cubical simulation box is replicated throughout 
space to form an infinite lattic).  Simulations were carried out both with and
without the use of symmetry adapted crossover and the population size
was 1000.  100 independent runs were carried out, and the mean value of
the best energies at each step is displayed in Fig.~\ref{prog2} along
with the runs with two seeds. For this value of $J$ the ground state
has an energy of -2600.

The superior performance of SAC compared to that of using a conventional
crossover is evident although comparison of the SAC results from
Figs.~\ref{prog1} and~\ref{prog2} shows that the evolution is markedly
slower when $J = 1$.  Populations with more than 1000 individuals were
required to find the global minima in 50\% of the runs, and on average,
200 000 evolutionary steps were needed.  To locate the global minimum
with 96\% probability large populations ($> 2500$ members) must be used.

The very different performance of the GA applied to the two Ising models
is not surprising: In the case $J = -1$ there are only 2 distinct
global minima, independent of system size, whereas for $J = 1$ there
are $3\cdot L$ distinct solutions, where $L$ is the side length of the
cube, i.e.~in our case there are 30 different solutions.  A synchronized
search towards one of the 2 global minima is thus much more rapid than
a synchronized search in the direction of the $3\cdot L$ global minima.
One way to overcome this slow convergence may be via the use of pinning
by e.g.~adding a energy penalty to a large number of global minima,
i.e. to break translation symmetry and thereby forcing the evolution in
the direction of only a few global minima.  Work along these lines is in
progress~\cite{Mohn:tbp}.

\section {Conclusion}

We have developed a real-space symmetry adapted genetic algorithm
for finding global minima configurations of alloys and disordered
ceramics. The method is illustrated using two simple model-systems
mimicking alloys that either phase separate or order at low temperature.
Conventional genetic algorithms are unable to solve these problems
because the inherent space group symmetry prevents a 
constant driving force in the direction of a single global minimum.
We avoid this synchronisation problem by incorporating the lattice symmetry of the
system within the GA operators by means of replacing the child with a randomly
chosen {\it symmetrically equivalent} child. Use of symmetry adaption
in combination with real-space crossover operator 
has the advantage of moving the local evolution of good ``clusters''
to other region in space where the evolution has halted, preventing 
premature convergence. The
advantage of using a real-space crossover in conjunction with SAC is
evident by  the comparison of results where a uniformed (binary) SAC were applied.
Our results are highly encouraging for the atomistic understanding of
more complex alloys and substitutionally disordered ceramics.

\section{Acknowledgement}

Computational facilities were made available though a grant of computing
time for the Program for supercomputing, Norway. W.K.~is a senior member
of the Institute Universitaire de France, whose support is gratefully
acknowledged.

\newpage

\newpage 

\newpage

\newpage

% \label{countor1}

\newpage

\newpage

\newpage

\end{document}